\documentstyle[preprint,aps]{revtex}
\begin{document}

\draft

\title{Spin injection efficiency from two adjoining ferromagnetic metals 
into a two-dimensional electron gas}

\author{Jun~Wang$^{1}$, D.~Y.~Xing$^{1}$, and H.~B.~Sun$^{2}$}
\address{$^1$National Laboratory of Solid State Microstructures and Department of Physics,\\
Nanjing University, Nanjing 210093, China}
\address{$^2$Department of Physics, University of Queensland, Brisbane Qld 4072, 
Australia}

\date{\today}
\maketitle

\begin{abstract}
In order to enhance spin injection efficiency from ferromagnetic (FM) metal into 
a two-dimensional electron gas (2DEG), we introduce another FM metal and two tunnel 
barriers (I) between them to investigate the current polarization in such 
ballistic FM/I/FM/I/2DEG junction. Our treatment is based on the free-electron 
scattering theory. It is found that due to quantum interference effect, the 
magnitude and sign of the current polarization exhibits periodical oscillating 
behavior with variation of the thickness of the middle FM metal layer or its 
exchange energy strength. For some suitable parameters, the spin injection 
efficiency may arrive over $80\%$ in this junction and can also be controlled 
by the electron density of 2DEG. Our results may shed light on the 
development of new spin-polarized device
\end{abstract}


In the recent years there have been much theoretical and experimental work
in the spin electronics (spintronics) field[1-3], in which the degrees of freedom 
of both electronic spin and charge are exploited. The magnetoelectronic device 
based on the spin-polarized transport in the semiconductors, which was first 
proposed by Datta and Das[4], has numerous potential applications in the 
information technology (IT) industry. The injection of spin-polarized carriers from 
ferromagnetic (FM) semiconductor into nonmagnetic semiconductor (SM)[5-6] has 
been achieved successfully with an efficiency $\sim90\%$. Jonker {\it{et al.}}[7]
even observed full polarized current by using an external magnetic field. 
Whereas spin injection from FM metal into SM is more attractive because
FM metals such as Fe have a relatively high Cuire temperature, which 
makes them indispensable for the room temperature devices. However, 
the spin injection efficiency in this FM/SM junction are very low and moreover,
there exist much debate on it[8].    

As Schmidt {\it{et al.}}[9] pointed out, the basic obstacle for spin-polarized 
injection from FM metal into SM in the diffusive system results from the 
conductivity mismatch between them. Although many authors[10-12] have shown that 
this kind of conductivity mismatch could be improved by introduction of a tunnel 
barrier (I) between them, which can assume the tunnel conductance difference 
between two spin channels, the efficiency of spin injection still remain low
in comparison with that from ferromagnetic-SM into SM. For instance,
by interposing a tunnel barrier between FM metal and SM, 
Zhu {\it{et al.}}[13] have observed experimentally $2\%$ efficiency of spin injection from 
Fe into n-GaAs at room temperature; Heersche {\it{et al.}}[14] theoretically calculated this
ballistic FM/I/2DEG(two-dimensional electron gas)[15] junction and obtained $\sim 10\%$ 
current polarization. A Schottky barrier formed at the Fe/AlGaAs interface by
 Hanbicki {\it{et al.}}[16] as a tunnel barrier can make the efficiency 
of spin injection $\sim 30\%$ in this junction. 

In the present work, we show theoretically that the high efficiency of spin 
injection from FM metal into 2DEG might be achieved by introducing another 
FM material (FM metal or ferromagnetic SM) between them besides two tunnel 
barriers. In the ballistic approximation, we treat this FM/I/FM/I/2DEG junction 
with the free-electron scattering theory, which has been widespread 
employed to deal with the interface scattering of electrons[17-18]. The first FM metal 
of the FM/I/FM/I/2DEG junction is a source of spin injection electrons (FM1), 
while the middle FM metal is taken as a resonant device to tune the tunnel current 
(FM2). Due to the quantum interference effect, the moderate thickness of the 
FM2 layer or its strength of spin exchange splitting energy may induce very
high degree of current polarization. FM1 can even be a normal metal in our model 
since FM2 is crucial to cause the spin-polarized current. Increasing exchange energy
of both FM1 and FM2 as well as the strength of two tunnel barriers 
would lead to enhancement of current polarization.  
The electron density of 2DEG affects the quantum interference effect so 
that it can also influence the degree of current polarization.

In the free electron approximation, the Hamiltonian for the FM/I/FM/I/2DEG junction 
reads
\begin{equation}
{\cal H}=\frac{-{\hbar}^{2}}{2m}{\Delta}^{2}+V(x)+U_{1}\delta(x)+U_{2}\delta(x-L)-
\theta(-x){\bf h_{1}}{\cdot}{\bf\sigma}-\theta(x)\theta(L-x)
{\bf h_{2}}{\cdot}{\bf {\sigma}},
\end{equation}
where $m$ is the effective electron mass, $m=m_{e}$ in two FM metals for $x<L$ and 
$m=m_{s}$ in 2DEG for $x>L$. Here we hypothesize that FM1 and FM2 have
same effective electron masses. $h_{1}$ and $h_{2}$ are respectively the internal 
molecular fields of the FM1 and FM2 layer and $\sigma$ denotes the Pauli spin operator. 
$\theta (x)$ is the step function. The two thin tunnel barriers are described by 
$\delta$-type potentials, which does not lose generality. We wish to point out
that even our two-dimensional model were replaced by three-dimensional one with
different barrier shape such as rectangle one, or Schottky barrier between FM and 
SM, the qualitative results in this papers would not change. $U_{1}$ at $x=0$ and 
$U_{2}$ at $x=L$ are related with the barrier's width and height. The potential energy $V(x)$ is 
zero for $x<L$ and $E_{B}$ for $x>L$. The schematic band structures of the FM/I/FM/I/2DEG 
junction is shown in Fig.~1. The spin quantum axis is taken along $y$ direction 
and the magnetizations of two FM metals are assumed to be parallel for simplicity  
while the net tunnel current flows in the $x$ direction.

In the two-band model, the energy eigenvalues of a single electron
with spin $\sigma$ ($\uparrow$ or $\downarrow$) are 
$E^{fm1}_{\uparrow}=(\hbar{\bf K}^{fm1}_{\uparrow})^{2}/2m_{e}$ and 
$E^{fm1}_{\downarrow}=(\hbar{\bf K}^{fm1}_{\downarrow})^{2}/2m_{e}+\Delta_{1}$ 
in the FM1 layer, $E^{fm2}_{\uparrow}=(\hbar{\bf K}^{fm2}_{\uparrow})^{2}/2m_{e}$
and $E^{fm2}_{\downarrow}=(\hbar{\bf K}^{fm2}_{\downarrow})^{2}/2m_{e}+\Delta_{2}$
in FM2 layer, and $E^{sm}_{\sigma}=E_{B}+
(\hbar{\bf K}^{sm}_{\sigma})^{2}/2m_{s}$ in 2DEG, where 
$\Delta_{1}=2h_{1}$ and $\Delta_{2}=2h_{2}$ are the exchange energies of
FM1 and FM2, respectively. $E_{B}$ is the difference between the lower 
conduction-band edge of 2DEG and that of FM. In the two-dimensional system,
the density of states in 2DEG is constant for the energy dispersion of free electrons
and $E^{sm}_{\sigma}=E_{B}+\pi{\hbar}^{2}n_{2DEG}/m_{s}$ with 
$n_{2DEG}$ being the electron density of 2DEG. In the small bias approximation,
only electrons near the Fermi energy ($E_{F}$) surface  contribute greatly to 
the net tunnel current so that
we can take $E_{\uparrow}^{fm1}=E^{fm2}_{\uparrow}
=E_{\uparrow}^{sm}=E_{F}$ and  $E_{\downarrow}^{fm1}=
E^{fm2}_{\downarrow}=E_{\downarrow}^{sm}=E_{F}$. Thus, the
magnitude of Fermi wave vectors in three regions can be explicitly expressed as
\begin{equation}
k^{fm1(2)}_{\uparrow}=\frac{1}{\hbar}\sqrt{2m_{e}E_{F}},
\end{equation}
\begin{equation}
k^{fm1(2)}_{\downarrow}=\frac{1}{\hbar}\sqrt{2m_{e}(E_{F}-\Delta_{1(2)})},
\end{equation}
and
\begin{equation}
k^{sm}_{\sigma}=\sqrt{2\pi n_{2DEG}}.
\end{equation}
It is assumed that the interfaces between the tunnel barriers and FM metals or 2DEG are
ideally smooth and without diffusive scattering so that the momentum along $y$ direction keep
constant when electrons are scattered by them. We define 
$k_{\sigma}^{y}=k^{fm1}_{\sigma}\cdot\sin\phi$
in three regions and the corresponding momenta along $x$ direction become
\begin{equation}
k^{fm1}_{\sigma,x}(\phi)=k^{fm1}_{\sigma}\cdot\cos\phi
\end{equation}
for $x<0$,
\begin{equation}
k^{fm2}_{\sigma,x}(\phi)=\sqrt{(k^{fm2}_{\sigma})^{2}-(k_{\sigma}^{y})^{2}}
\end{equation}
for $0<x<L$,
and
\begin{equation}
k^{sm}_{\sigma,x}(\phi)=\sqrt{(k^{sm}_{\sigma})^{2}-(k_{\sigma}^{y})^{2}}
\end{equation}  
for $L<x$.

Considering a single electron tunneling through the FM/I/FM/I/2DEG junction,
a reflective wave would appear in the FM1 layer and a transmission wave in 2DEG. 
In the FM2 layer, the electron will be multireflected due to the presence of 
two barriers. For some suitable width (L) of the FM2 layer, resonant reflection 
or transmission may occur. This may in turn result in high degree of spin injection 
from FM1 into 2DEG. The wave functions in three regions are given by
\begin{equation}
\psi_{\sigma}(fm1)=e^{(i{\bf k}^{fm1}_{\sigma}{\bf \cdot r})}+r_{\sigma}
e^{(-i{\bf k}^{fm1}_{\sigma}{\bf \cdot r})}
\end{equation}
for $x<0$,
\begin{equation}
\psi_{\sigma}(fm2)=a_{\sigma}e^{(i{\bf k}^{fm2}_{\sigma}{\bf \cdot r})}+b_{\sigma}
e^{(-i{\bf k}^{fm2}_{\sigma}{\bf \cdot r})}
\end{equation}
for $0<x<L$,
and
\begin{equation}
\psi_{\sigma}(sm)=t_{\sigma}e^{(i{\bf k}^{sm}_{\sigma}{\bf \cdot r})}
\end{equation}
for $x>L$. Here, $r_{\sigma}$, $a_{\sigma}$, $b_{\sigma}$, and $t_{\sigma}$ 
are spin-dependent parameters. According to the requirements of wave functions 
continuing and their derivatives continuing at scattering interface $x=0$ and $x=L$, 
the transmission amplitude $t_{\sigma}$ of a single electron tunneling through 
this FM/I/FM/I/2DEG junction is straightforward 
\begin{equation}
t_{\sigma}=\frac{\frac{\alpha^{++}}{\alpha^{-+}}-\frac{\alpha^{--}}{\alpha^{+-}}
}{\frac{\beta^{+}}{\alpha^{-+}}exp(-ik^{fm2}_{\sigma,x})-\frac{\beta^{-}}{\alpha^{+-}}
exp(ik^{fm2}_{\sigma,x})}.
\end{equation}
where $\alpha^{\pm\pm}=ik^{fm2}_{\sigma,x}\pm ik^{fm1}_{\sigma,x}\pm Q_{1}$ and 
$\beta^{\pm}=ik^{fm2}_{\sigma,x}\pm i(k^{sm}_{\sigma,x})'\mp Q_{2}$ with 
$(k^{sm}_{\sigma,x})'={m_{e}}k^{sm}_{\sigma,x}/m_{s}$ and $Q_{1(2)}=2m_{e}U_{1(2)}/\hbar^{2}$.
When the junction is applied on a small voltage $V$ and $K_{B}T\ll E_{F}$ for low 
temperature, the spin-dependent charge current density can be evaluated by[14,19]
\begin{equation}
J_{\sigma}=\frac{e^{2}Vk^{fm1}_{\sigma}}{h\pi}\int^{\phi^{C}_{\sigma}}_{0}{d\phi
T_{\sigma}\cos\sigma},
\end{equation}
where the transmission coefficient 
$T_{\sigma}=m_{e}k^{sm}_{\sigma,x}|t|^{2}/m_{s}k^{fm1}_{\sigma,x}$ and
$\phi_{\sigma}^{C}$ is the critical incident angle of electrons in FM1
 to guarantee all momenta appearing 
in the integral to be real variables, i.e, no attenuating wave occurs in 2DEG.
$\phi_{\sigma}^{C}$ is determined by Eq.(7) due to $k^{sm}_{\sigma}\ll k^{fm1}_{\sigma}$,
 $k^{fm2}_{\sigma}$.

From equations above, we can calculate numerically the spin-dependent current density as 
a function of the thickness (L) of the FM2 layer, in which multireflection would lead to
resonant transmission of the electronic wave. Thus the transmission coefficient $T_{\sigma}$ will 
exhibit oscillating behavior as well as the current density $J_{\sigma}$. Their periods can be
approximately expressed as $L_{\sigma}=\pi/k^{fm2}_{\sigma}$ since the critical angle  
$\phi_{\sigma}^{C}$ is very mall from Eq.~(7). Due to the presence of exchange energy of
FM2 and $k^{fm2}_{\uparrow}\neq k^{fm2}_{\downarrow}$, the current density 
$J_{\uparrow}$ and $J_{\downarrow}$ have different vibrating periods so that 
the current polarization $p=(J_{\uparrow}-J_{\downarrow})/(J_{\uparrow}+J_{\downarrow})$ 
may arrive rather high degree at some suitable thickness $L$ as shown in Fig.~2.
This characterization is the same as the tunnel magnetoresistance (TMR) effect 
in the FM/I/NM/I/FM junction[20] with NM denoting a normal metal, in which high TMR 
could be achieved because of the resonant transmission of electronic wave in the NM layer. 
In Fig.~2, the short and long periodical vibration of the current polarization $P$
results from the superposition of two different periods of $J_{\uparrow}$ and 
$J_{\downarrow}$. With the variation of L, plus maximum and minus maximum of $P$
would alternates to appear. From Fig.~(2A), even a normal metal ($\Delta_{1}=0$)
as FM1 can also lead to current polarization. This is just because the FM2 layer 
exists. Enlarging the exchange energy ($\Delta_{1}$) of FM1, the overall profile of 
$P$ keeps invariable whereas its amplitude increases much (the solid line 
{\bf{e}} in Fig.~(2B)), i.e., increasing the spin-polarized degree of FM1 
will raise the spin injection efficiency from FM metal 
into 2DEG. When $L=0$ and our model becomes FM/I/2DEG junction[14], the normal metal
($\Delta_{1}$=0) as FM1 would result in zero current polarizaiton (Fig.~2A). 
While another FM metal (FM2) is interposed into such a single junction, the current 
polarization ($L\neq 0$) would increase greatly  in comparison with that of $L=0$
as shown in Fig.~(2A-2B).
As is generally admitted, the tunnel barriers can 
result in the growth of the current polarization, which is indeed found in our 
calculation by comparing three lines in Fig.~(2A). The two tunnel barriers strength 
 increase  from $Z_{1(2)}=0.5$ to $Z_{1(2)}=1.0$ by two steps (the dimensionless 
$Z$ is defined as $Z_{1(2)}=Q_{1(2)}/k_{\uparrow}^{fm1}$), both of them
lead to a remarkable increase of the current polarization.

It is also shown in Fig.~(2B) that the electron density of 2DEG $n_{2DEG}$ affects 
the spin injection efficiency. The critical incident angle $\phi_{\sigma}^{C}$ 
of electrons in FM1 
is determined by $n_{2DEG}$ (Eq.~(7)). The larger $n_{2DEG}$ will 
widen $\phi_{\sigma}^{C}$
and the tunnel current density $J_{\sigma}$ increases greatly,
whereas the spin injection efficiency decreases. 
In this FM/I/FM/2DEG junction, the high degree of current polarization originates from 
the quantum interference effect, which usually had better be single moded in order to
obtain large effects  for different phase shift exist in different modes.  
The larger $\phi_{\sigma}^{C}$ will introduce the 
more modes and the quantum interference effect tend to wash out 
so that the current polarization 
could decrease. This is very interesting since the $n_{2DEG}$ can 
be easily controlled by an external gate voltage[21-22].

Although the maximum of the current polarization $P$ in Fig.~2 
might be obtained by tuning the thickness $L$ of FM2, it oscillates quickly
with increase of L and the spin-flipping 
length of electrons in FM metal is rather smaller than that in SM[23]. These may cause 
that the maximum of $P$ is difficult to be found in experiment. 
However, the FM2 layer in our model can lead to different oscillating periods of
$J_{\uparrow}$ and $J_{\uparrow}$. Thus, fixing its thickness $L$ and varying its 
exchange energy ($\Delta_{2}$), the vibrating behavior of current polarization $P$ should 
also appear. This is actually true as shown in Fig.~3. Here
$L$ is taken as $15$ ($\AA$) that is much less than the spin flipping length in FM metal. 
With increase of $\Delta_{2}$, $P$ will alternate its sign and it may take its maximum 
over $80\%$ at some suitable values of $\Delta_{2}$. Since the overall trend of 
minority spin current density $J_{\downarrow}$ decreases with increase of $\Delta_{2}$, 
the maximum of plus current polarization $P$ will keep upgoing. 
When $\Delta_{2}>E_{F}$, the FM2 would be a half-metal, i.e., 
it is a well for spin-up (majority spin) electrons 
and a rectangle barrier for spin-down (minority spin) electrons, 
then the high degree of current polarization would easily form since the transmission 
coefficient of spin-up electron is much larger than that of spin-down electron. 
This case has been discussed by Carlos Egues[24]. From Fig.~3, it is suggested that
the high degree of current polarization in FM/I/FM/2DEG junction 
should be achieved by tuning the magnitude 
of exchange energy ($\Delta_{2}$) of FM2 in experiment. For instance, we may apply an external 
magnetic field on FM2 and vary its strength or we would change the concentration 
of magnetic ions if FM2 was a ferromagnetic semiconductor.

In summary, we have investigated  the spin injection efficiency from 
two adjoining FM metals into 2DEG with two tunnel barriers among them
by employing the free-electron scattering theory. 
Making use of the quantum interference effect in this ballistic 
FM/I/FM/I/2DEG junction, the high degree
of current polarization may be achieved with some suitable thickness $L$ of the FM2 
layer or its exchange energy $\Delta_{2}$.  It is also shown 
that the electron density of 2DEG can alter the spin injection efficiency. Our model in this paper
would be an alternative to achieve high efficiency of spin injection from FM metal into 2DEG.

\newpage 

\begin{figure}[tbp]
\caption{Band structures of two adjoining FM metals and 2DEG. Two tunnel barriers lie
at $x=0$ and $x=L$. $\Delta_{1}$ and $\Delta_{2}$ are respectively the exchange energies
of two FM. $\uparrow$ and $\downarrow$ represent spin-up band and spin-down band.
}
\end{figure}

\begin{figure}[tbp]
\caption{The current polarization $P=\frac{J_{\uparrow}-J_{\downarrow}}
{J_{\uparrow}-J_{\downarrow}}$ as a function of the thickness $L$ of the FM2 layer.
The parameters are taken as $E_{F}=2.5$ eV, $\Delta_{2}/E_{F}=0.3$,
and $m_{s}/m_{e}=0.06$.  
(A) $\Delta_{1}/E_{F}=0$, $n_{2DEG}=3.0*10^{12}\text{cm}^{-2}$,
 solid line (a) $Z_{1}=Z_{2}=0.5$,  dot line (b) $Z_{1}=1.0$, $Z_{2}=0.5$, and 
 dash  line (c) $Z_{1}=Z_{2}=1.0$. (B) $\Delta_{1}/E_{F}=0.8$, $Z_{1}=Z_{2}=1.0$, 
 the solid line (e) $n_{2DEG}=3.0*10^{12}\text{cm}^{-2}$,
 and dot line (f) $n_{2DEG}=0.5*10^{12}\text{cm}^{-2}$.
}
\end{figure}

\begin{figure}[tbp]
\caption{The current polarization $P$ as a function of the exchange energy strength 
 $\Delta_{2}/E_{F}$ of FM2.
$n_{2DEG}=3.0*10^{12}\text{cm}^{-2}$ and $L=15 \AA$. Other parameters are taken as 
$\Delta_{1}/E_{F}=0.$, $Z_{1}=Z_{2}=0.5$ for the solid line (a),  
$\Delta_{1}/E_{F}=0.$, $Z_{1}=Z_{2}=1.0$ for the dot line (b),
and $\Delta_{1}/E_{F}=0.8$, $Z_{1}=Z_{2}=1.0$ for the dash dot line (c).
}
\end{figure}

\end{document}